\documentclass[conference]{IEEEtran}
\IEEEoverridecommandlockouts

\usepackage{cite}
\usepackage{amsmath,amssymb,amsfonts}
\usepackage{amsthm}
\newtheorem{proposition}{Proposition}
\usepackage{algorithmic}
\usepackage{graphicx}
\usepackage{textcomp}
\usepackage{xcolor}
\usepackage{algorithm}
\newcommand{\STATEX}{\item[]}

\usepackage{booktabs}
\usepackage{array}
\usepackage{multirow}
\usepackage{makecell}
\usepackage{nicematrix}
\usepackage{arydshln}
\usepackage[caption=false,font=footnotesize]{subfig} 
\usepackage{stfloats}
\usepackage{url}
\usepackage{verbatim}
\usepackage{bm}
\usepackage{fp}
\usepackage{xfp}

\def\BibTeX{{\rm B\kern-.05em{\sc i\kern-.025em b}\kern-.08em
    T\kern-.1667em\lower.7ex\hbox{E}\kern-.125emX}}
\begin{document}

\title{MACS: Measurement-Aware Consistency Sampling for Inverse Problems
}

\author{\IEEEauthorblockN{Amirreza Tanevardi}
\IEEEauthorblockA{\textit{Department of Electrical Engineering} \\
\textit{Sharif University of Technology}\\
Tehran, Iran \\
amirreza.tanevardi13@sharif.edu}
\and
\IEEEauthorblockN{Pooria Abbas Rad Moghadam}
\IEEEauthorblockA{\textit{Department of Electrical Engineering} \\
\textit{Sharif University of Technology}\\
Tehran, Iran \\
p.abbasrad@ee.sharif.edu}
\and
\IEEEauthorblockN{Seyed Mohammad Eshtehardian}
\IEEEauthorblockA{\textit{Department of Electrical Engineering} \\
\textit{Sharif University of Technology}\\
Tehran, Iran \\
mohammad.eshtehardian@sharif.edu}
\and
\IEEEauthorblockN{Sajjad Amini}
\IEEEauthorblockA{\textit{Electronics Research Institute} \\
\textit{Sharif University of Technology}\\
Tehran, Iran \\
s\_amini@sharif.edu}
\and
\IEEEauthorblockN{Babak Khalaj}
\IEEEauthorblockA{\textit{Department of Electrical Engineering} \\
\textit{Sharif University of Technology}\\
Tehran, Iran \\
khalaj@sharif.edu}
\and
}

\maketitle

\begin{abstract}
Diffusion models have emerged as powerful generative priors for solving inverse imaging problems. However, their practical deployment is hindered by the substantial computational cost of slow, multi-step sampling. Although Consistency Models (CMs) address this limitation by enabling high-quality generation in only one or a few steps, their direct application to inverse problems has remained largely unexplored. This paper introduces a modified consistency sampling framework specifically designed for inverse problems. The proposed approach regulates the sampler's stochasticity through a measurement-consistency mechanism that leverages the degradation operator, thereby enforcing fidelity to the observed data while preserving the computational efficiency of consistency-based generation.
Comprehensive experiments on the Fashion-MNIST and LSUN Bedroom datasets demonstrate consistent improvements across both perceptual and pixel-level metrics, including the Fréchet Inception Distance (FID), Kernel Inception Distance (KID), peak signal-to-noise ratio (PSNR), and structural similarity index measure (SSIM), compared with baseline consistency and diffusion-based sampling methods. The proposed method achieves competitive or superior reconstruction quality with only a small number of sampling steps.
\end{abstract}


\section{Introduction}
\IEEEPARstart{I}{nverse} problems represent a fundamental class of challenges in computational imaging and signal processing, with critical applications in medical imaging~\cite{illposed}, remote sensing~\cite{comsenspriors}, seismic analysis~\cite{seismic}, video reconstruction~\cite{ho2022video,yeh2024diffir2vr,daras2024warped,kwon2024vision,kwon2025solving}, and audio signal processing~\cite{audiosig}. The core objective is to reconstruct the original signal $\boldsymbol{x}$ from degraded measurements $\boldsymbol{y}$ acquired through a degradation operator $\mathcal{A}$, which is typically ill-conditioned or non-invertible. This inherent ill-posedness necessitates the incorporation of strong prior knowledge to obtain meaningful reconstructions.

Traditional inverse problem methods rely on filtering, optimization~\cite{dabov2007image,elad2006image,gu2014weighted}, and classical machine learning techniques~\cite{elad2023image,gurrola2021residual,chen2016trainable,ohayon2021high}. Recently, diffusion models~\cite{DDPM,song2019generative,dhariwal2021diffusion,rombach2022high,song2020score} have become powerful priors for inverse problems, achieving state-of-the-art performance across tasks such as super-resolution~\cite{sup-res,saharia2022image}, inpainting~\cite{inpaint,lugmayr2022repaint}, deblurring~\cite{deblur}, and colorization~\cite{saharia2022palette}, while motivating latent-space methods~\cite{song2023solving,raphaeli2025silo} and many diffusion-based solvers~\cite{chung2022improving,chung2023dps,chung2024decomposed,chung2023solving3d,kawar2021snips,song2023pseudoinverse,wang2022zero,ILVR,postinv}. However, their reliance on hundreds to thousands of NFEs remains a major limitation for practical deployment~\cite{DDIM}.

Consistency Models (CMs)~\cite{song2023consistency,multistepCM,geng2024consistency} address this limitation by enabling high-quality generation in only one or a few sampling steps. However, unlike diffusion models, CMs lack an inherent mechanism for enforcing measurement consistency during sampling, making efficient CM-based inverse solvers an open research problem.

To address this challenge, we propose \emph{Measurement-Aware Consistency Sampling} (MACS), an aDDIM-based~\cite{multistepCM} test-time sampling framework that incorporates adaptive residual-based guidance into CM sampling. By regulating sampling stochasticity according to the measurement residual, MACS improves measurement consistency while preserving the efficiency of CMs, without requiring retraining or modification of the pretrained backbone.

We evaluate MACS on the Fashion-MNIST~\cite{fashion} and LSUN Bedroom~\cite{lsun} datasets across super-resolution, inpainting, and deblurring tasks. The proposed method consistently improves perceptual quality (FID~\cite{fid}, KID~\cite{kid}) while maintaining competitive reconstruction fidelity (PSNR and SSIM~\cite{ssim}) using only a few sampling steps.

The main contributions of this work are as follows:
\begin{itemize}
\item We propose a measurement-aware sampling framework for CMs that incorporates an adaptive residual-based guidance mechanism, which regulates sampling stochasticity to enhance fidelity while preserving few-step efficiency.

\item We provide extensive experimental results across multiple inverse imaging tasks, demonstrating that the proposed framework consistently improves reconstruction quality with only a small number of sampling steps.
\end{itemize}

\section{Preliminaries}

\subsection{Problem Formulation}

Inverse problems aim to recover an unknown signal $\boldsymbol{x} \in \mathbb{R}^n$ from a set of observed measurements $\boldsymbol{y} \in \mathbb{R}^m$, generated via a forward process:
\begin{equation}
    \boldsymbol{y} = \mathcal{A}(\boldsymbol{x}) + \boldsymbol{n},
    \label{eq:inverse_prob}
\end{equation}
where $\mathcal{A}: \mathbb{R}^n \rightarrow \mathbb{R}^m$ denotes the forward operator (e.g., the Radon transform in computed tomography or nonlinear compression), and $\boldsymbol{n}$ represents measurement noise. The problem is technically ill-posed when $m < n$ or when the operator is singular, implying that multiple solutions $\boldsymbol{x}$ may map to the same observation $\boldsymbol{y}$~\cite{illposed}.

\subsection{Diffusion Models}

Diffusion models learn the underlying data distribution by progressively corrupting clean samples with Gaussian noise and then reversing this process using a learned denoising network~\cite{DDPM,song2020score}. The forward diffusion process is defined as a Markov chain
\begin{equation}
    q(\boldsymbol{x}_t|\boldsymbol{x}_{t-1}) =
    \mathcal{N}(\boldsymbol{x}_t;
    \sqrt{1-\beta_t}\boldsymbol{x}_{t-1},
    \beta_t\boldsymbol{I}),
\end{equation}
where $\{\beta_t\}_{t=1}^{T}$ is a predefined variance schedule. By defining $\alpha_t=1-\beta_t$ and $\bar{\alpha}_t=\prod_{s=1}^{t}\alpha_s$, the noisy sample at timestep $t$ can be obtained directly as
\begin{equation}
    q(\boldsymbol{x}_t|\boldsymbol{x}_0)=
    \mathcal{N}(\boldsymbol{x}_t;
    \sqrt{\bar{\alpha}_t}\boldsymbol{x}_0,
    (1-\bar{\alpha}_t)\boldsymbol{I}).
\end{equation}

Starting from Gaussian noise, the reverse diffusion process iteratively reconstructs a clean sample using a neural network $\boldsymbol{\epsilon}_{\boldsymbol{\theta}}$ that predicts the injected noise:
\begin{equation}
    \boldsymbol{x}_{t-1}=
    \frac{1}{\sqrt{\alpha_t}}
    \left(
    \boldsymbol{x}_t-
    \frac{1-\alpha_t}{\sqrt{1-\bar{\alpha}_t}}
    \boldsymbol{\epsilon}_{\boldsymbol{\theta}}(\boldsymbol{x}_t,t)
    \right)
    +\tilde{\beta}_t\boldsymbol{\epsilon},
\end{equation}
where $\boldsymbol{\epsilon}\sim\mathcal{N}(0,\boldsymbol{I})$ and $\tilde{\beta}_t=\frac{1-\bar{\alpha}_{t-1}}{1-\bar{\alpha}_t}\beta_t$. The denoising network is trained to predict the added Gaussian noise using a mean squared error objective.

\subsection{Consistency Models}

Consistency Models (CMs)~\cite{song2023consistency,LCM} accelerate generative modeling by learning a direct mapping from any point on the probability flow ordinary differential equation (ODE) trajectory to the corresponding clean sample, enabling one- or few-step generation. Unlike diffusion models, which require iterative denoising, a CM learns a self-consistent mapping satisfying
\begin{equation}
f_\theta(\boldsymbol{x}_t,t)=f_\theta(\boldsymbol{x}_{t'},t'),
\qquad \forall\, t,t'\in[\epsilon,T],
\end{equation}
together with the boundary condition
\begin{equation}
f_\theta(\boldsymbol{x}_\epsilon,\epsilon)=\boldsymbol{x}_\epsilon.
\end{equation}
To satisfy this constraint, the model is parameterized as
\begin{equation}
f_\theta(\boldsymbol{x},t)
=
c_{\mathrm{skip}}(t)\boldsymbol{x}
+
c_{\mathrm{out}}(t)F_\theta(\boldsymbol{x},t),
\end{equation}
where the coefficients are designed such that $c_{\mathrm{skip}}(\epsilon)=1$ and $c_{\mathrm{out}}(\epsilon)=0$.

CMs support both single-step and few-step sampling. In the single-step setting, a clean sample is obtained directly from Gaussian noise,
\begin{equation}
\hat{\boldsymbol{x}}_0=f_\theta(\boldsymbol{x}_T,T),
\end{equation}
while the few-step variant alternates between consistency prediction and controlled noise injection to progressively refine the result~\cite{multistepCM}. By requiring only 1--4 network evaluations, CMs achieve a favorable balance between computational efficiency and sample quality, making them particularly attractive for inverse problems~\cite{song2023consistency}.

\subsection{DDIM}

Denoising Diffusion Implicit Models (DDIM)~\cite{DDIM} accelerate diffusion sampling by replacing the stochastic reverse process with a non-Markovian, often deterministic trajectory, enabling high-quality generation with significantly fewer sampling steps. Given a noisy sample $\boldsymbol{x}_t$, DDIM first estimates the corresponding clean image as
\begin{equation}
\hat{\boldsymbol{x}}_0(\boldsymbol{x}_t,t)
=
\frac{1}{\sqrt{\bar{\alpha}_t}}
\left(
\boldsymbol{x}_t
-
\sqrt{1-\bar{\alpha}_t}\,
\boldsymbol{\epsilon}_\theta(\boldsymbol{x}_t,t)
\right),
\end{equation}
where $\boldsymbol{\epsilon}_\theta$ denotes the pretrained noise prediction network.

Using Tweedie's formula and rewriting the formulation under variance-exploding parameterization, the DDIM update can be expressed in the compact affine form
\begin{equation}
\boldsymbol{x}_s^{\mathrm{DDIM}}
=
(1-\sqrt{\rho})\,\hat{\boldsymbol{x}}_0
+
\sqrt{\rho}\,\boldsymbol{x}_t,
\label{eq:ddim-linear}
\end{equation}
where $s = t_{i-1}$ denotes the next noise level, $t = t_i$ represents the current one and
\begin{equation}
\rho
=
\frac{\sigma_s^2}{\sigma_t^2}
=
\frac{s^2-t_{\min}^2}{t^2-t_{\min}^2}
\end{equation}
is the normalized noise-level ratio. This formulation forms the basis of many accelerated diffusion and consistency-based sampling methods by providing an efficient transition between consecutive noise levels with substantially fewer network evaluations.

\subsection{Adjusted DDIM Sampler}

The Adjusted DDIM (aDDIM) sampler~\cite{multistepCM} extends deterministic DDIM sampling by incorporating variance compensation to address the systematic underestimation of norms in standard DDIM, which degrades reconstruction quality in few-step sampling~\cite{multistepCM}.

The update rule in aDDIM is defined as
\begin{align}
    &\boldsymbol{x}_{s} = \hat{\boldsymbol{x}} +
    \sqrt{\rho + \left(\frac{1 - \sqrt{\rho}}{\|\hat{\boldsymbol{\epsilon}}\|}\right)^2 x_{\mathrm{var},t}} \, \hat{\boldsymbol{\epsilon}}
\end{align}
where $x_{\mathrm{var},t} = \eta \|\boldsymbol{x} - \hat{\boldsymbol{x}}\|^2$, $\eta$ is a tunable hyperparameter, and $\boldsymbol{x}$ is a \textbf{teacher} signal from either the dataset or a target diffusion model.

This formulation accounts for the discrepancy between DDIM trajectories and the true conditional distribution. Ideally, a clean sample $\boldsymbol{x}^* \sim p(\boldsymbol{x} \mid \boldsymbol{x}_t)$ preserves the proper distribution via:
\begin{equation}
    \boldsymbol{x}_s^* = (1 - \sqrt{\rho}) \boldsymbol{x}^* + \sqrt{\rho} \boldsymbol{x}_t.
\end{equation}
However, DDIM replaces $\boldsymbol{x}^*$ with the conditional expectation $\hat{\boldsymbol{x}} = \mathbb{E}[\boldsymbol{x} \mid \boldsymbol{x}_t]$. The resulting discrepancy in their squared norms is:
\begin{align}
    \mathbb{E}\!\left[\|\boldsymbol{x}_s^*\|^2 - \|\boldsymbol{x}_s^{\mathrm{DDIM}}\|^2 \mid \boldsymbol{x}_t\right]
    = \mathbb{E}\!\left[\|\boldsymbol{x}_s\|^2 \mid \boldsymbol{x}_t\right] - \|\boldsymbol{x}_s^{\mathrm{DDIM}}\|^2.
\end{align}
Since $\boldsymbol{x}_s^{\mathrm{DDIM}} = \mathbb{E}[\boldsymbol{x}_s \mid \boldsymbol{x}_t]$, this simplifies to the omitted variance:
\begin{align}
    \mathbb{E}\!\left[\|\boldsymbol{x}_s\|^2 \mid \boldsymbol{x}_t\right] - \left\|\mathbb{E}[\boldsymbol{x}_s \mid \boldsymbol{x}_t]\right\|^2
    = \mathrm{trace}\!\left(\mathrm{Var}[\boldsymbol{x}_s \mid \boldsymbol{x}_t]\right).
\end{align}
This omitted term can be expressed as:
$
   (1 - \sqrt{\rho})^2 \, \mathrm{trace}\!\left(\mathrm{Var}[\boldsymbol{x} \mid \boldsymbol{x}_t]\right)
$
where $\mathrm{trace}\!\left(\mathrm{Var}[\boldsymbol{x} \mid \boldsymbol{x}_t]\right)$ is approximated by $\eta \|\boldsymbol{x} - \hat{\boldsymbol{x}}\|^2$. By neglecting $\mathrm{Var}[\boldsymbol{x}_s \mid \boldsymbol{x}_t]$, standard DDIM produces smaller norms and over-smoothed results. aDDIM corrects this by scaling the added noise estimate by the conditional variance, ensuring accurate trajectory guidance for few-step inverse problems.

\section{Related Work}
\label{sec:related_work}

This section reviews recent advancements in generative modeling for inverse problems. We first discuss diffusion-based solvers, distinguishing between supervised and zero-shot strategies, and then examine the emerging application of CMs in this domain.

\subsection{Diffusion-Based Inverse Solvers}

Recently, diffusion models have emerged as powerful priors for inverse problems. Existing methods are broadly divided into supervised approaches, which retrain or fine-tune diffusion models for specific tasks~\cite{saharia2022palette,elata2025invfussion}, and zero-shot approaches, which use pretrained unconditional models to sample from the posterior distribution. In zero-shot methods, the posterior score is decomposed into the pretrained prior score and an approximated likelihood score:
\begin{equation}
\nabla_{\boldsymbol{x}_t}\log p(\boldsymbol{x}_t|\boldsymbol{y})
=
\nabla_{\boldsymbol{x}_t}\log p(\boldsymbol{x}_t)
+
\nabla_{\boldsymbol{x}_t}\log p(\boldsymbol{y}|\boldsymbol{x}_t),
\label{eq:bayes_score}
\end{equation}
Most zero-shot methods differ in how they estimate the likelihood gradient, including DPS~\cite{chung2023dps}, LGD~\cite{song2023loss}, and DEFT~\cite{denker2024deft}, while DDRM~\cite{ddrm} instead performs posterior sampling in the spectral domain for linear inverse problems. Comprehensive surveys are available in~\cite{daras2024survey}.

\subsection{Consistency Models for Inverse Problems}

Consistency Models (CMs) offer a computationally efficient alternative to diffusion models by enabling one- or few-step generation, but their limited sampling trajectories make enforcing measurement consistency more challenging. Recent approaches address this through different conditioning mechanisms, including auxiliary measurement encoders in CoSIGN~\cite{cosign} and SBI~\cite{sbi}, latent-space and prompt-optimization strategies in LATINO and LATINO-PRO~\cite{spagnoletti2025latino}, and the zero-shot CM4IR framework~\cite{garber2025cm4ir}, which combines pseudo-inverse initialization, back-projection guidance, and noise injection. These methods highlight the growing potential of CMs for efficient inverse problem solving.

\section{Method}

Sampling from CMs for inverse problems requires balancing measurement fidelity with generative quality. Existing approaches often rely on auxiliary encoders~\cite{cosign,sbi} or text-conditioned priors and multiple passes through encoders and decoders~\cite{spagnoletti2025latino}, thereby overlooking the potential of the sampling process itself to improve reconstruction quality. Our objective is to develop a sampling strategy that enhances reconstruction quality while preserving the fast sampling characteristics of CMs~\cite{song2023consistency,multistepCM}.

We propose the \emph{Measurement-Aware Consistency Sampler (MACS)}, a measurement-aware consistency sampling method that extends the aDDIM framework by replacing its variance term with a \emph{residual} term. This modification enables adaptive stochasticity control guided by measurement fidelity.

The MACS update can be expressed as
\begin{align}
    &\hat{\boldsymbol{x}} = f_{\theta}(\boldsymbol{x}_t, \boldsymbol{y}, t), 
    \qquad \hat{\boldsymbol{\epsilon}} = \boldsymbol{x}_t - \hat{\boldsymbol{x}} \\
    &\boldsymbol{x}_{s}^{\text{MACS}} = \hat{\boldsymbol{x}} +
    \sqrt{\rho + \left(\frac{1 - \sqrt{\rho}}{\|\hat{\boldsymbol{\epsilon}}\|}\right)^2 
    \gamma \|\boldsymbol{y} - \mathcal{A}(\hat{\boldsymbol{x}})\|^2 } 
    \, \hat{\boldsymbol{\epsilon}} 
\end{align}

\begin{proposition}[Justification of the Residual Substitution]
\label{prop:residual-substitution}
Assume the measurement model
\begin{align}
\boldsymbol{y}
&=
\mathcal{A}(\boldsymbol{x})+\boldsymbol{n},
\end{align}
where $\mathcal{A}$ is linear and
$\boldsymbol{n}\sim\mathcal{N}(\boldsymbol{0},\sigma_y^2\boldsymbol{I})$.
Then, for any reconstruction estimate $\hat{\boldsymbol{x}}$,
\begin{align}
\mathbb{E}\!\left[\|\boldsymbol{y}-\mathcal{A}(\hat{\boldsymbol{x}})\|^2\right]
&=
\|\mathcal{A}(\boldsymbol{x}-\hat{\boldsymbol{x}})\|^2
+
m\sigma_y^2
\notag\\
&\le
\|\mathcal{A}\|_2^2
\|\boldsymbol{x}-\hat{\boldsymbol{x}}\|^2
+
m\sigma_y^2.
\end{align}

Consequently, up to an additive constant and a scaling factor, minimizing the measurement residual is equivalent to minimizing the image-domain reconstruction error.
\end{proposition}

\begin{proof}
By linearity of $\mathcal{A}$, the measurement residual decomposes as
\begin{align}
\boldsymbol{y}-\mathcal{A}(\hat{\boldsymbol{x}})
&=
\mathcal{A}(\boldsymbol{x}-\hat{\boldsymbol{x}})
+\boldsymbol{n}.
\end{align}

Expanding the squared norm and taking expectations, while using the independence of $\boldsymbol{n}$ from $(\boldsymbol{x},\hat{\boldsymbol{x}})$ and $\mathbb{E}[\boldsymbol{n}]=\boldsymbol{0}$, yields
\begin{align}
\mathbb{E}\!\left[\|\boldsymbol{y}-\mathcal{A}(\hat{\boldsymbol{x}})\|^2\right]
&=
\mathbb{E}\!\left[
\|\mathcal{A}(\boldsymbol{x}-\hat{\boldsymbol{x}})+\boldsymbol{n}\|^2
\right]
\\
&=
\|\mathcal{A}(\boldsymbol{x}-\hat{\boldsymbol{x}})\|^2
+
\mathbb{E}\!\left[\|\boldsymbol{n}\|^2\right]
\\
&=
\|\mathcal{A}(\boldsymbol{x}-\hat{\boldsymbol{x}})\|^2
+
\mathrm{trace}(\Sigma_n),
\end{align}
where $\Sigma_n=\sigma_y^2\boldsymbol{I}$ and the cross term vanishes. Since
$\mathrm{trace}(\Sigma_n)=m\sigma_y^2$, we obtain
\begin{align}
\mathbb{E}\!\left[\|\boldsymbol{y}-\mathcal{A}(\hat{\boldsymbol{x}})\|^2\right]
&=
\|\mathcal{A}(\boldsymbol{x}-\hat{\boldsymbol{x}})\|^2
+
m\sigma_y^2.
\end{align}

Applying the spectral-norm bound
$\|\mathcal{A}\boldsymbol{u}\|^2 \le \|\mathcal{A}\|_2^2 \|\boldsymbol{u}\|^2$
with $\boldsymbol{u}=\boldsymbol{x}-\hat{\boldsymbol{x}}$ gives the stated inequality.
Because $m\sigma_y^2$ is independent of $\hat{\boldsymbol{x}}$, minimizing the expected measurement residual is equivalent to minimizing $\|\mathcal{A}(\boldsymbol{x}-\hat{\boldsymbol{x}})\|^2$, and hence, up to the factor $\|\mathcal{A}\|_2^2$, to minimizing the image-domain error $\|\boldsymbol{x}-\hat{\boldsymbol{x}}\|^2$.
When the noise floor $m\sigma_y^2$ is negligible in practice, it can be absorbed---together with $\|\mathcal{A}\|_2^2$---into a tunable hyperparameter $\eta'$, so that
\begin{align}
\mathbb{E}\!\left[\|\boldsymbol{y}-\mathcal{A}(\hat{\boldsymbol{x}})\|^2\right]
&\le
\eta'\,\|\boldsymbol{x}-\hat{\boldsymbol{x}}\|^2.
\end{align}

Thus, with a suitable choice of $\gamma$, the residual term used in MACS remains fully consistent with the variance-compensation mechanism of aDDIM, while naturally extending it to measurement-driven reconstruction tasks.
\end{proof}

This formulation generalizes aDDIM by introducing a data-dependent correction term that reflects how closely the current estimate aligns with the measurements.

\begingroup
\begin{algorithm}[!t]
    \caption{MACS: Measurement-Aware Consistency Sampling for Inverse Reconstruction}
    \label{alg:macsampler}
    \begin{algorithmic}[1]
        \STATEX \textbf{Input:} measurements $\boldsymbol{y}$, consistency model $f_\theta$, forward operator $\mathcal{A}$, guidance scale $\gamma$, noise levels $\{t_i\}_{i=0}^N$, minimal noise level $t_{\text{min}}$
        \newline
        \STATEX \textbf{Output:} reconstruction $\hat{\boldsymbol{x}}_0$
        \STATE Sample $\boldsymbol{x}_{t_N} \sim \mathcal{N}(0, I)$
        \FOR{$i = N$ down to $2$}
            \STATE $t \leftarrow t_i$, $s \leftarrow t_{i-1}$
            \STATE $\hat{\boldsymbol{x}} \leftarrow f_\theta(\boldsymbol{x}_t, \boldsymbol{y}, t)$
            \STATE $\hat{\boldsymbol{\epsilon}} \leftarrow \boldsymbol{x}_t - \hat{\boldsymbol{x}}$
            \STATE $r \leftarrow \gamma \|\boldsymbol{y} - \mathcal{A}(\hat{\boldsymbol{x}})\|^2$
            \vspace{2pt}
            \STATE $\rho \leftarrow \dfrac{s^2 - t_{\text{min}}^2}{t^2 - t_{\text{min}}^2}$
            \STATE $\boldsymbol{x}_s^{\text{MACS}} \leftarrow \hat{\boldsymbol{x}} + \sqrt{\rho + r\left(\dfrac{1 - \sqrt{\rho}}{\|\hat{\boldsymbol{\epsilon}}\|}\right)^2}\,\hat{\boldsymbol{\epsilon}}$
        \ENDFOR
        \STATE $\hat{\boldsymbol{x}}_0 \leftarrow f_\theta(\boldsymbol{x}_{t_1}^{\text{MACS}}, \boldsymbol{y}, t_1)$
        \RETURN $\hat{\boldsymbol{x}}_0$
    \end{algorithmic}
\end{algorithm}
\endgroup

\begin{figure}[!t]
    \centering
    \includegraphics[width=\columnwidth]{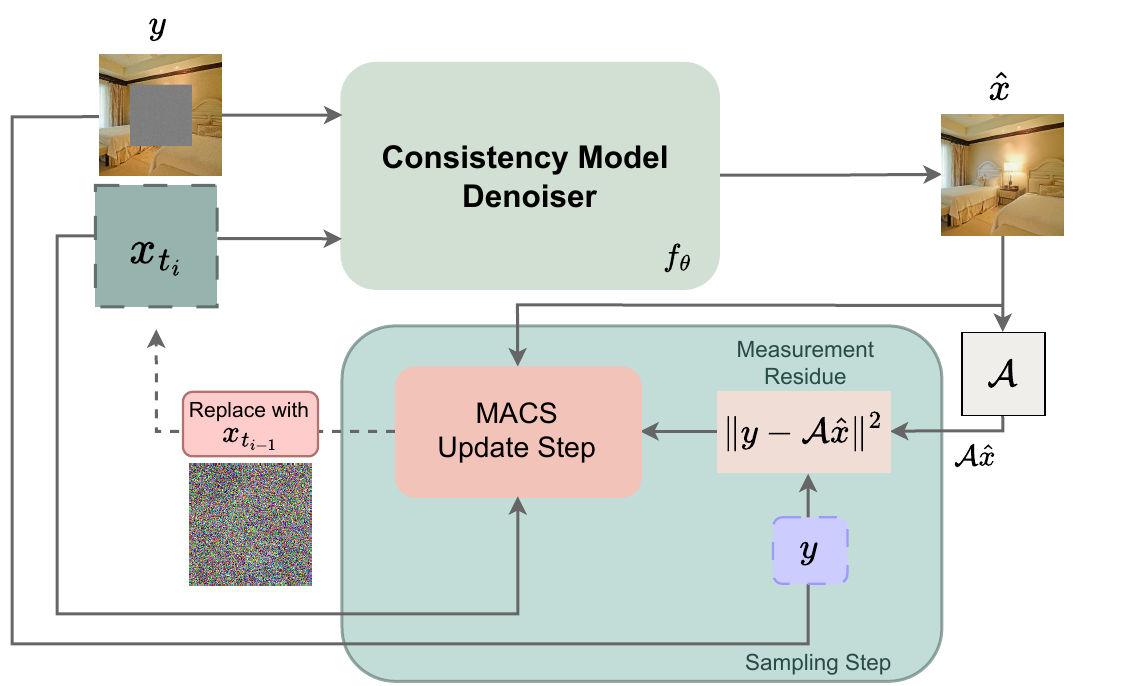}
    \caption{Illustration of the proposed MACS framework. The diagram shows the iterative sampling process in which measurement consistency is enforced through the residual term. The CM $f_\theta$ denoises the noisy latent $\boldsymbol{x}_t$ conditioned on the measurements $\boldsymbol{y}$, and the MACS update adaptively adjusts the stochasticity based on this residual.}
    \label{fig:illustration}
\end{figure}

Intuitively, the higher the residual, the more freedom the CM denoiser has over the generation of details, leading to finer reconstructions due to the greater magnitude of added noise.

Fig.~\ref{fig:illustration} depicts the general CM sampling trajectory under MACS, where the new noisy latent obtained through the proposed measurement-aware update is iteratively refined by the CM $f_\theta$ in a few sampling steps.

In addition to the scalar hyperparameter $\gamma$, MACS allows for further design choices, such as step-dependent schedules for $\gamma$ and potential normalization of the residual by an estimate of $\|\mathcal{A}\|_2^2$ or the measurement noise level $\sigma_y$. In this work, however, we focus on a simple constant-$\gamma$ configuration that already yields consistent gains. The full procedure is outlined in Algorithm~\ref{alg:macsampler}.

\section{Experiments}
\label{sec:experiments}

To validate the proposed algorithm, we conduct extensive experiments on multiple inverse problems using two representative consistency-model frameworks. Our evaluation covers super-resolution, inpainting, and deblurring tasks across both high-resolution natural images and grayscale image reconstruction. We compare MACS with several strong few-step samplers while keeping the pretrained consistency models unchanged, demonstrating that the observed improvements arise solely from the proposed sampling strategy.

\begingroup
\def\totalwidth{1}
\def\numdegredations{3}
\def\nummetrics{4}
\FPeval{\totalcolumns}{round(\numdegredations*\nummetrics +1 ,0)}
\FPeval{\colwidth}{\totalwidth/\totalcolumns}

\newcommand{\rankhl}[4]{%
  \ifnum\numexpr\pdfstrcmp{#1}{#2}=0
    \textbf{#1}
  \else\ifnum\numexpr\pdfstrcmp{#1}{#3}=0
    \underline{#1}
  \else
    #1
  \fi\fi
}

\newcommand{\rankhlrev}[4]{%
  \ifnum\numexpr\pdfstrcmp{#1}{#4}=0
    \textbf{#1}
  \else\ifnum\numexpr\pdfstrcmp{#1}{#3}=0
    \underline{#1}
  \else
    #1
  \fi\fi
}

\edef\firstTime{148}\edef\secondTime{149}\edef\thirdTime{331}
\edef\firstPSNRa{26.03}\edef\secondPSNRa{25.70}\edef\thirdPSNRa{25.51}
\edef\firstSSIMa{0.771}\edef\secondSSIMa{0.768}\edef\thirdSSIMa{0.765}
\edef\firstFIDa{40.90}\edef\secondFIDa{40.85}\edef\thirdFIDa{40.13}
\edef\firstKIDa{5.83}\edef\secondKIDa{5.67}\edef\thirdKIDa{5.54}

\edef\firstPSNRb{22.51}\edef\secondPSNRb{22.40}\edef\thirdPSNRb{22.11}
\edef\firstSSIMb{0.830}\edef\secondSSIMb{0.825}\edef\thirdSSIMb{0.779}
\edef\firstFIDb{39.92}\edef\secondFIDb{39.82}\edef\thirdFIDb{39.44}
\edef\firstKIDb{4.72}\edef\secondKIDb{4.69}\edef\thirdKIDb{4.54}

\edef\firstPSNRc{24.41}\edef\secondPSNRc{24.02}\edef\thirdPSNRc{23.09}
\edef\firstSSIMc{0.700}\edef\secondSSIMc{0.690}\edef\thirdSSIMc{0.672}
\edef\firstFIDc{41.57}\edef\secondFIDc{41.56}\edef\thirdFIDc{40.90}
\edef\firstKIDc{9.32}\edef\secondKIDc{8.61}\edef\thirdKIDc{8.32}

\begin{table*}[!t]
    \centering
    \setlength{\tabcolsep}{4pt}
    \renewcommand{\arraystretch}{1.15}
    \caption{Performance comparison of Baseline and MACS across three inverse tasks on the LSUN Bedroom dataset.
    Bold and underlined numbers denote the best and the second-best results, respectively.}
    \begin{tabular}{l *{\totalcolumns}{c}}
        \hline
        \multicolumn{2}{c}{} &
        \multicolumn{\nummetrics}{c}{Super-Resolution $\times4$} &
        \multicolumn{\nummetrics}{c}{Inpainting} &
        \multicolumn{\nummetrics}{c}{Linear Deblurring}
        \\
        \cmidrule(lr){3-6}
        \cmidrule(lr){7-10}
        \cmidrule(lr){11-14}
        Method &
        NFE &
        PSNR & SSIM & FID & KID &
        PSNR & SSIM & FID & KID &
        PSNR & SSIM & FID & KID \\
        \hline
        DPM-Solver~\cite{dpm-solver}& 
        \rankhl{8}{\firstTime}{\secondTime}{\thirdTime}&
        \rankhl{18.22}{\firstPSNRa}{\secondPSNRa}{\thirdPSNRa}&
        \rankhl{0.338}{\firstSSIMa}{\secondSSIMa}{\thirdSSIMa}&
        \rankhlrev{75.22}{\firstFIDa}{\secondFIDa}{\thirdFIDa}&
        \rankhlrev{25.30}{\firstKIDa}{\secondKIDa}{\thirdKIDa}&
        \rankhl{19.32}{\firstPSNRb}{\secondPSNRb}{\thirdPSNRb}&
        \rankhl{0.759}{\firstSSIMb}{\secondSSIMb}{\thirdSSIMb}&
        \rankhlrev{48.09}{\firstFIDb}{\secondFIDb}{\thirdFIDb}&
        \rankhlrev{5.48}{\firstKIDb}{\secondKIDb}{\thirdKIDb} &
        \rankhl{16.97}{\firstPSNRc}{\secondPSNRc}{\thirdPSNRc}&
        \rankhl{0.288}{\firstSSIMc}{\secondSSIMc}{\thirdSSIMc}&
        \rankhlrev{81.94}{\firstFIDc}{\secondFIDc}{\thirdFIDc}&
        \rankhlrev{33.10}{\firstKIDc}{\secondKIDc}{\thirdKIDc}
        \\
        Heun~\cite{kerras2022} & 
        \rankhl{40}{\firstTime}{\secondTime}{\thirdTime}&
        \rankhl{17.57}{\firstPSNRa}{\secondPSNRa}{\thirdPSNRa}&
        \rankhl{0.270}{\firstSSIMa}{\secondSSIMa}{\thirdSSIMa}&
        \rankhlrev{84.67}{\firstFIDa}{\secondFIDa}{\thirdFIDa}&
        \rankhlrev{31.8}{\firstKIDa}{\secondKIDa}{\thirdKIDa}&
        \rankhl{19.53}{\firstPSNRb}{\secondPSNRb}{\thirdPSNRb}&
        \rankhl{0.779}{\firstSSIMb}{\secondSSIMb}{\thirdSSIMb}&
        \rankhlrev{45.86}{\firstFIDb}{\secondFIDb}{\thirdFIDb}&
        \rankhlrev{4.92}{\firstKIDb}{\secondKIDb}{\thirdKIDb}&
        \rankhl{16.54}{\firstPSNRc}{\secondPSNRc}{\thirdPSNRc}&
        \rankhl{0.219}{\firstSSIMc}{\secondSSIMc}{\thirdSSIMc}&
        \rankhlrev{98.04}{\firstFIDc}{\secondFIDc}{\thirdFIDc}&
        \rankhlrev{46.8}{\firstKIDc}{\secondKIDc}{\thirdKIDc}
        \\
        Euler~\cite{kerras2022}& 
        \rankhl{2}{\firstTime}{\secondTime}{\thirdTime}&
        \rankhl{25.51}{\firstPSNRa}{\secondPSNRa}{\thirdPSNRa}&
        \rankhl{0.768}{\firstSSIMa}{\secondSSIMa}{\thirdSSIMa}&
        \rankhlrev{40.85}{\firstFIDa}{\secondFIDa}{\thirdFIDa}&
        \rankhlrev{5.83}{\firstKIDa}{\secondKIDa}{\thirdKIDa}&
        \rankhl{22.40}{\firstPSNRb}{\secondPSNRb}{\thirdPSNRb}&
        \rankhl{0.825}{\firstSSIMb}{\secondSSIMb}{\thirdSSIMb}&
        \rankhlrev{39.92}{\firstFIDb}{\secondFIDb}{\thirdFIDb}&
        \rankhlrev{5.09}{\firstKIDb}{\secondKIDb}{\thirdKIDb}&
        \rankhl{23.09}{\firstPSNRc}{\secondPSNRc}{\thirdPSNRc}&
        \rankhl{0.672}{\firstSSIMc}{\secondSSIMc}{\thirdSSIMc}&
        \rankhlrev{41.56}{\firstFIDc}{\secondFIDc}{\thirdFIDc}&
        \rankhlrev{9.32}{\firstKIDc}{\secondKIDc}{\thirdKIDc}
        \\
        Multistep~\cite{song2023consistency}& 
        \rankhl{2}{\firstTime}{\secondTime}{\thirdTime}&
        \rankhl{26.03}{\firstPSNRa}{\secondPSNRa}{\thirdPSNRa}&
        \rankhl{0.771}{\firstSSIMa}{\secondSSIMa}{\thirdSSIMa}&
        \rankhlrev{40.90}{\firstFIDa}{\secondFIDa}{\thirdFIDa}&
        \rankhlrev{5.67}{\firstKIDa}{\secondKIDa}{\thirdKIDa}&
        \rankhl{22.51}{\firstPSNRb}{\secondPSNRb}{\thirdPSNRb}&
        \rankhl{0.830}{\firstSSIMb}{\secondSSIMb}{\thirdSSIMb}&
        \rankhlrev{39.82}{\firstFIDb}{\secondFIDb}{\thirdFIDb}&
        \rankhlrev{4.69}{\firstKIDb}{\secondKIDb}{\thirdKIDb} &
        \rankhl{24.41}{\firstPSNRc}{\secondPSNRc}{\thirdPSNRc}&
        \rankhl{0.700}{\firstSSIMc}{\secondSSIMc}{\thirdSSIMc}&
        \rankhlrev{41.57}{\firstFIDc}{\secondFIDc}{\thirdFIDc}&
        \rankhlrev{8.61}{\firstKIDc}{\secondKIDc}{\thirdKIDc}
        \\
        MACS (Ours)& 
        \rankhl{2}{\firstTime}{\secondTime}{\thirdTime}&
        \rankhl{25.70}{\firstPSNRa}{\secondPSNRa}{\thirdPSNRa}&
        \rankhl{0.765}{\firstSSIMa}{\secondSSIMa}{\thirdSSIMa}&
        \rankhlrev{40.13}{\firstFIDa}{\secondFIDa}{\thirdFIDa}&
        \rankhlrev{5.54}{\firstKIDa}{\secondKIDa}{\thirdKIDa}&
        \rankhl{22.11}{\firstPSNRb}{\secondPSNRb}{\thirdPSNRb}&
        \rankhl{0.830}{\firstSSIMb}{\secondSSIMb}{\thirdSSIMb}&
        \rankhlrev{39.44}{\firstFIDb}{\secondFIDb}{\thirdFIDb}&
        \rankhlrev{4.54}{\firstKIDb}{\secondKIDb}{\thirdKIDb} &
        \rankhl{24.02}{\firstPSNRc}{\secondPSNRc}{\thirdPSNRc}&
        \rankhl{0.690}{\firstSSIMc}{\secondSSIMc}{\thirdSSIMc}&
        \rankhlrev{40.90}{\firstFIDc}{\secondFIDc}{\thirdFIDc}&
        \rankhlrev{8.32}{\firstKIDc}{\secondKIDc}{\thirdKIDc}
        \\
        \hline
    \end{tabular}
    \label{tab:results-LSUN-bedroom}
\end{table*}
\endgroup

\begingroup
\def\totalwidth{1}
\def\numdegredations{3}
\def\nummetrics{4}
\FPeval{\totalcolumns}{round(\numdegredations*\nummetrics +1 ,0)}
\FPeval{\colwidth}{\totalwidth/\totalcolumns}

\newcommand{\rankhl}[4]{%
  \ifnum\numexpr\pdfstrcmp{#1}{#2}=0
    \textbf{#1}
  \else\ifnum\numexpr\pdfstrcmp{#1}{#3}=0
    \underline{#1}
  \else
    #1
  \fi\fi
}

\newcommand{\rankhlrev}[4]{%
  \ifnum\numexpr\pdfstrcmp{#1}{#4}=0
    \textbf{#1}
  \else\ifnum\numexpr\pdfstrcmp{#1}{#3}=0
    \underline{#1}
  \else
    #1
  \fi\fi
}

\edef\firstTime{148}\edef\secondTime{149}\edef\thirdTime{331}
\edef\firstPSNRa{19.05}\edef\secondPSNRa{18.97}\edef\thirdPSNRa{18.76}
\edef\firstSSIMa{0.599}\edef\secondSSIMa{0.557}\edef\thirdSSIMa{0.518}
\edef\firstFIDa{22.34}\edef\secondFIDa{20.27}\edef\thirdFIDa{19.85}
\edef\firstKIDa{14.2}\edef\secondKIDa{13.3}\edef\thirdKIDa{12.1}

\edef\firstPSNRb{17.99}\edef\secondPSNRb{17.75}\edef\thirdPSNRb{16.59}
\edef\firstSSIMb{0.569}\edef\secondSSIMb{0.523}\edef\thirdSSIMb{0.516}
\edef\firstFIDb{19.84}\edef\secondFIDb{17.64}\edef\thirdFIDb{17.52}
\edef\firstKIDb{11.5}\edef\secondKIDb{10.3}\edef\thirdKIDb{9.6}

\edef\firstPSNRc{18.55}\edef\secondPSNRc{18.22}\edef\thirdPSNRc{18.09}
\edef\firstSSIMc{0.541}\edef\secondSSIMc{0.509}\edef\thirdSSIMc{0.509}
\edef\firstFIDc{23.60}\edef\secondFIDc{23.09}\edef\thirdFIDc{19.58}
\edef\firstKIDc{15.6}\edef\secondKIDc{15.1}\edef\thirdKIDc{12.9}

\begin{table*}[!t]
    \centering
    \setlength{\tabcolsep}{4pt}
    \renewcommand{\arraystretch}{1.15}
    \caption{Performance comparison of Baseline and MACS across three inverse tasks on the Fashion-MNIST dataset.
    Bold and underlined numbers denote the best and the second-best results, respectively.}
    \begin{tabular}{l *{\totalcolumns}{c}}
    \hline
        \multicolumn{2}{c}{} &
        \multicolumn{\nummetrics}{c}{Super-Resolution $\times2$} &
        \multicolumn{\nummetrics}{c}{Inpainting} &
        \multicolumn{\nummetrics}{c}{Linear Deblurring}
        \\
        \cmidrule(lr){3-6}
        \cmidrule(lr){7-10}
        \cmidrule(lr){11-14}
        Method &
        NFE &
        PSNR & SSIM & FID & KID &
        PSNR & SSIM & FID & KID &
        PSNR & SSIM & FID & KID \\
        \hline
        DPM-Solver~\cite{dpm-solver}& 
        \rankhl{8}{\firstTime}{\secondTime}{\thirdTime}&
        \rankhl{15.74}{\firstPSNRa}{\secondPSNRa}{\thirdPSNRa}&
        \rankhl{0.394}{\firstSSIMa}{\secondSSIMa}{\thirdSSIMa}&
        \rankhlrev{43.89}{\firstFIDa}{\secondFIDa}{\thirdFIDa}&
        \rankhlrev{31.4}{\firstKIDa}{\secondKIDa}{\thirdKIDa}&
        \rankhl{13.83}{\firstPSNRb}{\secondPSNRb}{\thirdPSNRb}&
        \rankhl{0.320}{\firstSSIMb}{\secondSSIMb}{\thirdSSIMb}&
        \rankhlrev{49.46}{\firstFIDb}{\secondFIDb}{\thirdFIDb}&
        \rankhlrev{36.5}{\firstKIDb}{\secondKIDb}{\thirdKIDb} &
        \rankhl{14.69}{\firstPSNRc}{\secondPSNRc}{\thirdPSNRc}&
        \rankhl{0.343}{\firstSSIMc}{\secondSSIMc}{\thirdSSIMc}&
        \rankhlrev{71.10}{\firstFIDc}{\secondFIDc}{\thirdFIDc}&
        \rankhlrev{55.2}{\firstKIDc}{\secondKIDc}{\thirdKIDc}
        \\
        Heun~\cite{kerras2022} & 
        \rankhl{40}{\firstTime}{\secondTime}{\thirdTime}&
        \rankhl{15.20}{\firstPSNRa}{\secondPSNRa}{\thirdPSNRa}&
        \rankhl{0.365}{\firstSSIMa}{\secondSSIMa}{\thirdSSIMa}&
        \rankhlrev{55.18}{\firstFIDa}{\secondFIDa}{\thirdFIDa}&
        \rankhlrev{41.2}{\firstKIDa}{\secondKIDa}{\thirdKIDa}&
        \rankhl{13.35}{\firstPSNRb}{\secondPSNRb}{\thirdPSNRb}&
        \rankhl{0.294}{\firstSSIMb}{\secondSSIMb}{\thirdSSIMb}&
        \rankhlrev{61.19}{\firstFIDb}{\secondFIDb}{\thirdFIDb}&
        \rankhlrev{47.1}{\firstKIDb}{\secondKIDb}{\thirdKIDb}&
        \rankhl{13.94}{\firstPSNRc}{\secondPSNRc}{\thirdPSNRc}&
        \rankhl{0.300}{\firstSSIMc}{\secondSSIMc}{\thirdSSIMc}&
        \rankhlrev{94.13}{\firstFIDc}{\secondFIDc}{\thirdFIDc}&
        \rankhlrev{78.7}{\firstKIDc}{\secondKIDc}{\thirdKIDc}
        \\
        Euler~\cite{kerras2022}& 
        \rankhl{2}{\firstTime}{\secondTime}{\thirdTime}&
        \rankhl{19.05}{\firstPSNRa}{\secondPSNRa}{\thirdPSNRa}&
        \rankhl{0.599}{\firstSSIMa}{\secondSSIMa}{\thirdSSIMa}&
        \rankhlrev{20.27}{\firstFIDa}{\secondFIDa}{\thirdFIDa}&
        \rankhlrev{13.3}{\firstKIDa}{\secondKIDa}{\thirdKIDa}&
        \rankhl{16.59}{\firstPSNRb}{\secondPSNRb}{\thirdPSNRb}&
        \rankhl{0.516}{\firstSSIMb}{\secondSSIMb}{\thirdSSIMb}&
        \rankhlrev{17.64}{\firstFIDb}{\secondFIDb}{\thirdFIDb}&
        \rankhlrev{10.3}{\firstKIDb}{\secondKIDb}{\thirdKIDb}&
        \rankhl{18.09}{\firstPSNRc}{\secondPSNRc}{\thirdPSNRc}&
        \rankhl{0.509}{\firstSSIMc}{\secondSSIMc}{\thirdSSIMc}&
        \rankhlrev{23.60}{\firstFIDc}{\secondFIDc}{\thirdFIDc}&
        \rankhlrev{15.1}{\firstKIDc}{\secondKIDc}{\thirdKIDc}
        \\
        Multistep~\cite{song2023consistency}& 
        \rankhl{2}{\firstTime}{\secondTime}{\thirdTime}&
        \rankhl{18.76}{\firstPSNRa}{\secondPSNRa}{\thirdPSNRa}&
        \rankhl{0.518}{\firstSSIMa}{\secondSSIMa}{\thirdSSIMa}&
        \rankhlrev{22.34}{\firstFIDa}{\secondFIDa}{\thirdFIDa}&
        \rankhlrev{14.2}{\firstKIDa}{\secondKIDa}{\thirdKIDa}&
        \rankhl{17.75}{\firstPSNRb}{\secondPSNRb}{\thirdPSNRb}&
        \rankhl{0.523}{\firstSSIMb}{\secondSSIMb}{\thirdSSIMb}&
        \rankhlrev{19.84}{\firstFIDb}{\secondFIDb}{\thirdFIDb}&
        \rankhlrev{11.5}{\firstKIDb}{\secondKIDb}{\thirdKIDb} &
        \rankhl{18.22}{\firstPSNRc}{\secondPSNRc}{\thirdPSNRc}&
        \rankhl{0.509}{\firstSSIMc}{\secondSSIMc}{\thirdSSIMc}&
        \rankhlrev{23.09}{\firstFIDc}{\secondFIDc}{\thirdFIDc}&
        \rankhlrev{15.6}{\firstKIDc}{\secondKIDc}{\thirdKIDc}
        \\
        MACS (Ours)& 
        \rankhl{2}{\firstTime}{\secondTime}{\thirdTime}&
        \rankhl{18.97}{\firstPSNRa}{\secondPSNRa}{\thirdPSNRa}&
        \rankhl{0.557}{\firstSSIMa}{\secondSSIMa}{\thirdSSIMa}&
        \rankhlrev{19.85}{\firstFIDa}{\secondFIDa}{\thirdFIDa}&
        \rankhlrev{12.1}{\firstKIDa}{\secondKIDa}{\thirdKIDa}&
        \rankhl{17.99}{\firstPSNRb}{\secondPSNRb}{\thirdPSNRb}&
        \rankhl{0.569}{\firstSSIMb}{\secondSSIMb}{\thirdSSIMb}&
        \rankhlrev{17.52}{\firstFIDb}{\secondFIDb}{\thirdFIDb}&
        \rankhlrev{9.6}{\firstKIDb}{\secondKIDb}{\thirdKIDb} &
        \rankhl{18.55}{\firstPSNRc}{\secondPSNRc}{\thirdPSNRc}&
        \rankhl{0.541}{\firstSSIMc}{\secondSSIMc}{\thirdSSIMc}&
        \rankhlrev{19.58}{\firstFIDc}{\secondFIDc}{\thirdFIDc}&
        \rankhlrev{12.9}{\firstKIDc}{\secondKIDc}{\thirdKIDc}
        \\
        \hline
    \end{tabular}
    \label{tab:results-Fashion-MNIST}
\end{table*}
\endgroup

\begingroup
\def\totalwidth{1}
\def\numdegredations{3}
\def\nummetrics{4}
\FPeval{\totalcolumns}{round(\numdegredations*\nummetrics +1 ,0)}
\FPeval{\colwidth}{\totalwidth/\totalcolumns}

\newcommand{\rankhl}[4]{%
  \ifnum\numexpr\pdfstrcmp{#1}{#2}=0
    \textbf{#1}
  \else\ifnum\numexpr\pdfstrcmp{#1}{#3}=0
    \underline{#1}
  \else
    #1
  \fi\fi
}

\newcommand{\rankhlrev}[4]{%
  \ifnum\numexpr\pdfstrcmp{#1}{#4}=0
    \textbf{#1}
  \else\ifnum\numexpr\pdfstrcmp{#1}{#3}=0
    \underline{#1}
  \else
    #1
  \fi\fi
}

\edef\firstTime{148}\edef\secondTime{149}\edef\thirdTime{331}
\edef\firstPSNRa{25.85}\edef\secondPSNRa{25.81}\edef\thirdPSNRa{22.47}
\edef\firstSSIMa{0.815}\edef\secondSSIMa{0.811}\edef\thirdSSIMa{0.791}
\edef\firstFIDa{40.70}\edef\secondFIDa{40.44}\edef\thirdFIDa{39.90}
\edef\firstKIDa{3.44}\edef\secondKIDa{3.10}\edef\thirdKIDa{3.07}

\begin{table}[!t]
    \centering
    \setlength{\tabcolsep}{4pt}
    \renewcommand{\arraystretch}{1.15}
    \caption{Performance comparison of Baseline and MACS for the Nonlinear Deblurring task on the LSUN Bedroom dataset.
    Bold and underlined numbers denote the best and the second-best results, respectively.}
    \begin{tabular}{l c c c c c}
    \hline
        \multicolumn{2}{c}{} &
        \multicolumn{4}{c}{Nonlinear Deblurring} \\
        \cmidrule(lr){3-6}
        Method &
        NFE&
        PSNR &
        SSIM &
        FID &
        KID \\
        \hline
        
        DPM-Solver~\cite{dpm-solver}& 
        \rankhl{8}{\firstTime}{\secondTime}{\thirdTime}&
        \rankhl{22.41}{\firstPSNRa}{\secondPSNRa}{\thirdPSNRa}&
        \rankhl{0.565}{\firstSSIMa}{\secondSSIMa}{\thirdSSIMa}&
        \rankhlrev{57.07}{\firstFIDa}{\secondFIDa}{\thirdFIDa}&
        \rankhlrev{12.5}{\firstKIDa}{\secondKIDa}{\thirdKIDa}
        \\
        Heun~\cite{kerras2022} & 
        \rankhl{40}{\firstTime}{\secondTime}{\thirdTime}&
        \rankhl{22.47}{\firstPSNRa}{\secondPSNRa}{\thirdPSNRa}&
        \rankhl{0.563}{\firstSSIMa}{\secondSSIMa}{\thirdSSIMa}&
        \rankhlrev{56.05}{\firstFIDa}{\secondFIDa}{\thirdFIDa}&
        \rankhlrev{13.4}{\firstKIDa}{\secondKIDa}{\thirdKIDa}
        \\
        Euler~\cite{kerras2022}& 
        \rankhl{2}{\firstTime}{\secondTime}{\thirdTime}&
        \rankhl{25.81}{\firstPSNRa}{\secondPSNRa}{\thirdPSNRa}&
        \rankhl{0.791}{\firstSSIMa}{\secondSSIMa}{\thirdSSIMa}&
        \rankhlrev{40.44}{\firstFIDa}{\secondFIDa}{\thirdFIDa}&
        \rankhlrev{3.44}{\firstKIDa}{\secondKIDa}{\thirdKIDa}
        \\
        Multistep~\cite{song2023consistency}& 
        \rankhl{2}{\firstTime}{\secondTime}{\thirdTime}&
        \rankhl{25.85}{\firstPSNRa}{\secondPSNRa}{\thirdPSNRa}&
        \rankhl{0.815}{\firstSSIMa}{\secondSSIMa}{\thirdSSIMa}&
        \rankhlrev{40.70}{\firstFIDa}{\secondFIDa}{\thirdFIDa}&
        \rankhlrev{3.10}{\firstKIDa}{\secondKIDa}{\thirdKIDa}
        \\
        MACS (Ours)& 
        \rankhl{2}{\firstTime}{\secondTime}{\thirdTime}&
        \rankhl{25.81}{\firstPSNRa}{\secondPSNRa}{\thirdPSNRa}&
        \rankhl{0.811}{\firstSSIMa}{\secondSSIMa}{\thirdSSIMa}&
        \rankhlrev{39.90}{\firstFIDa}{\secondFIDa}{\thirdFIDa}&
        \rankhlrev{3.07}{\firstKIDa}{\secondKIDa}{\thirdKIDa}
        \\
        \hline
    \end{tabular}
    \label{tab:results-NonlinearDeblur}
\end{table}
\endgroup

\begin{figure*}[!t]
\centering

\newcommand{\colw}{0.14\linewidth}

\newcommand{\roww}{0.14\linewidth} 

\newcommand{\coltitle}[1]{%
  \begin{minipage}[t]{\colw}
    \centering
    \small \textbf{#1}
  \end{minipage}%
}

\newcommand{\rowtitle}[1]{%
  \begin{minipage}[c]{\roww}
    \raggedleft
    \small #1
  \end{minipage}%
}

\newcommand{\subfig}[1]{%
  \begin{minipage}[t]{\colw}
    \includegraphics[width=\linewidth]{#1}
  \end{minipage}%
}

\coltitle{GT}%
\coltitle{Measurement}%
\coltitle{\footnotesize MACS (Ours)-2 steps}%
\coltitle{\footnotesize DPM-Solver-8 steps}%
\coltitle{Heun-40 steps}%
\coltitle{Euler-2 steps}%
\coltitle{Multistep-2 steps}\\[2pt]

\subfig{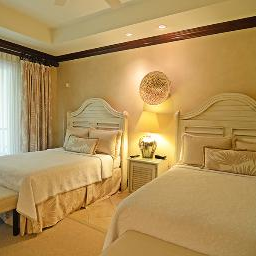}%
\subfig{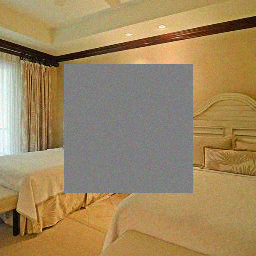}%
\subfig{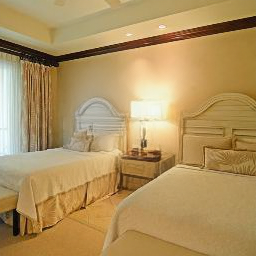}%
\subfig{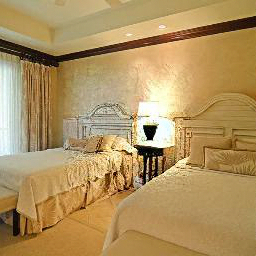}%
\subfig{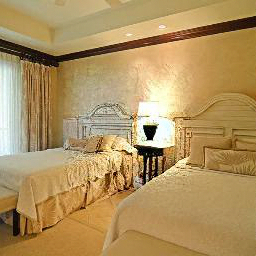}%
\subfig{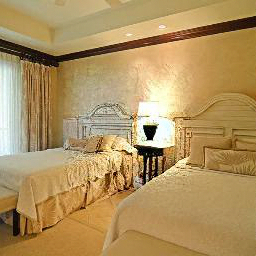}%
\subfig{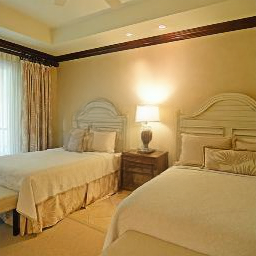}\\[3pt]

\subfig{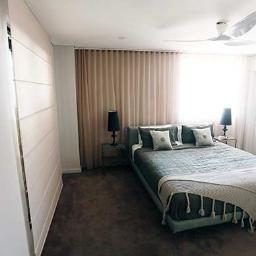}%
\subfig{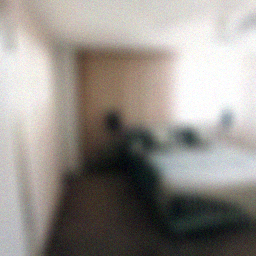}%
\subfig{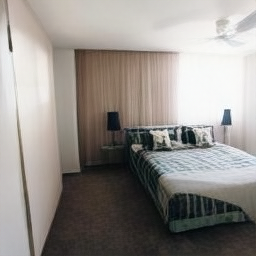}%
\subfig{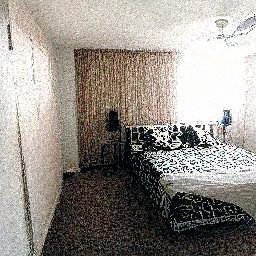}%
\subfig{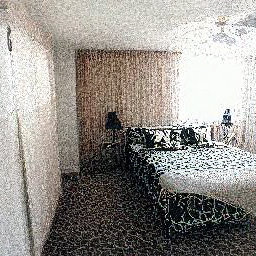}%
\subfig{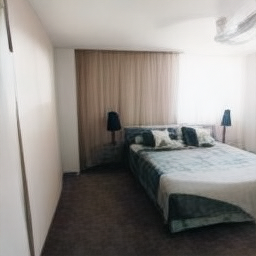}%
\subfig{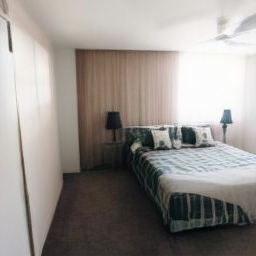}\\[3pt]

\subfig{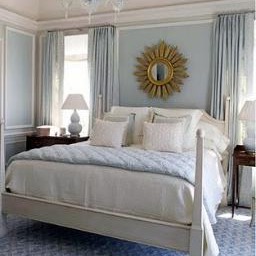}%
\subfig{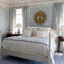}%
\subfig{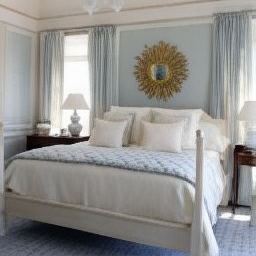}%
\subfig{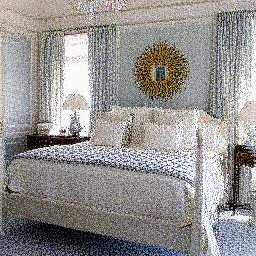}%
\subfig{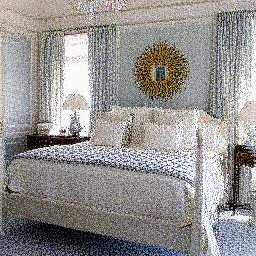}%
\subfig{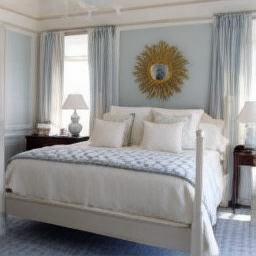}%
\subfig{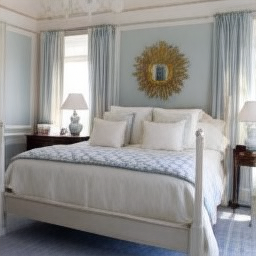}\\[3pt]

\subfig{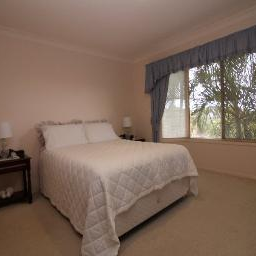}%
\subfig{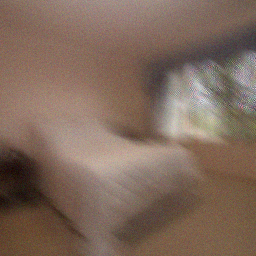}%
\subfig{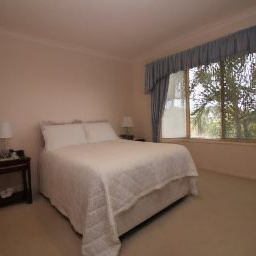}%
\subfig{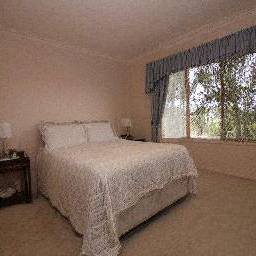}%
\subfig{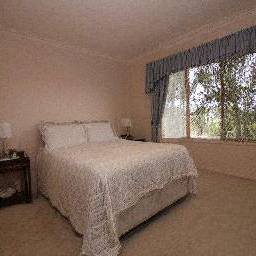}%
\subfig{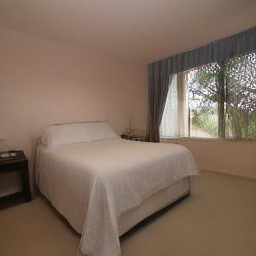}%
\subfig{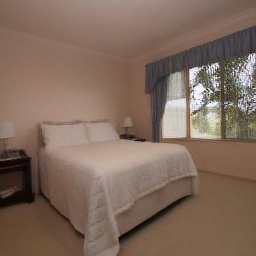}\\[3pt]

\caption{
Visual comparison across four inverse problems on the LSUN Bedroom dataset: inpainting (top row), linear deblurring (second row), super-resolution (third row), and nonlinear deblurring (bottom row). Each row shows the ground truth, measurement, our MACS reconstruction, and outputs from four fast-sampling baselines (DPM-Solver, Heun, Euler, and Multistep). MACS consistently reconstructs sharper edges and more details compared to other fast samplers.
}

\label{fig:comparison-lsun}
\end{figure*}

\subsection{Experimental Setup}

To evaluate the proposed method, we conduct experiments on the LSUN Bedroom dataset~\cite{lsun} at $256\times256$ resolution, containing over $3$ million training images and $300$ validation images, and the Fashion-MNIST dataset~\cite{fashion} at $28\times28$ resolution, containing over $60{,}000$ training images and $10{,}000$ validation images. For LSUN Bedroom, we use the CoSIGN base model~\cite{cosign}, which employs a trainable ControlNet while keeping the CM backbone frozen. We use the released checkpoints for inpainting, super-resolution, and nonlinear deblurring, and train a separate ControlNet for linear deblurring, which is not provided in~\cite{cosign}. For Fashion-MNIST, we adopt the SBI model~\cite{sbi}, where an auxiliary encoder conditions the CM using a learned measurement representation.

We evaluate four inverse problems. For super-resolution, LSUN Bedroom images are downsampled by a factor of $4$ in each spatial dimension, while Fashion-MNIST images are downsampled by a factor of $2$. For inpainting, a central square mask with half the image side length is applied. For linear deblurring, Gaussian blur with standard deviation $5$ is used for LSUN Bedroom and $3$ for Fashion-MNIST. Following~\cite{cosign}, Gaussian measurement noise with standard deviation $\sigma_y=0.05$ is added to all measurements. We further evaluate nonlinear deblurring on LSUN Bedroom using a learned nonlinear kernel model that produces spatially varying, image-dependent blur patterns.

Across all experiments, the CM backbones and conditioning networks remain fixed, and only the sampling strategy is changed, ensuring that performance differences are solely attributable to the proposed sampler. To guarantee reproducibility, identical noise schedules, ControlNet parameters, and preprocessing pipelines are used for all evaluated samplers.

\begin{figure*}[!t]
\centering

\newcommand{\colw}{0.14\linewidth}

\newcommand{\roww}{0.14\linewidth} 

\newcommand{\coltitle}[1]{%
  \begin{minipage}[t]{\colw}
    \centering
    \small \textbf{#1}
  \end{minipage}%
}

\newcommand{\rowtitle}[1]{%
  \begin{minipage}[c]{\roww}
    \raggedleft
    \small #1
  \end{minipage}%
}

\newcommand{\subfig}[1]{%
  \begin{minipage}[t]{\colw}
    \includegraphics[width=\linewidth]{#1}
  \end{minipage}%
}


\coltitle{Ground Truth}%
\coltitle{Measurement}%
\coltitle{\footnotesize MACS (Ours)-2 steps}%
\coltitle{\footnotesize DPM-Solver-8 steps}%
\coltitle{Heun-40 steps}%
\coltitle{Euler-2 steps}%
\coltitle{Multistep-2 steps}\\[2pt]

\subfig{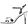}%
\subfig{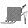}%
\subfig{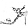}%
\subfig{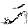}%
\subfig{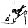}%
\subfig{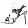}%
\subfig{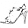}\\[3pt]

\subfig{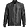}%
\subfig{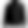}%
\subfig{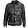}%
\subfig{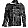}%
\subfig{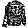}%
\subfig{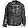}%
\subfig{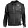}\\[3pt]

\subfig{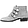}%
\subfig{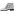}%
\subfig{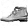}%
\subfig{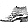}%
\subfig{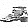}%
\subfig{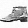}%
\subfig{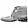}\\[3pt]

\caption{
Visual comparison on three inverse problems from the Fashion-MNIST dataset: inpainting (top), linear deblurring (middle), and super-resolution (bottom). Each row shows the ground truth, the measurement, our MACS reconstruction, and the outputs of four fast-sampling baselines. MACS yields clearer structures and more details than the other methods.
}

\label{fig:comparison-Fashion-mnist}

\end{figure*}

\begin{figure*}[!t]
\centering

\subfloat[LSUN Bedroom]{
    \includegraphics[width=0.48\textwidth]{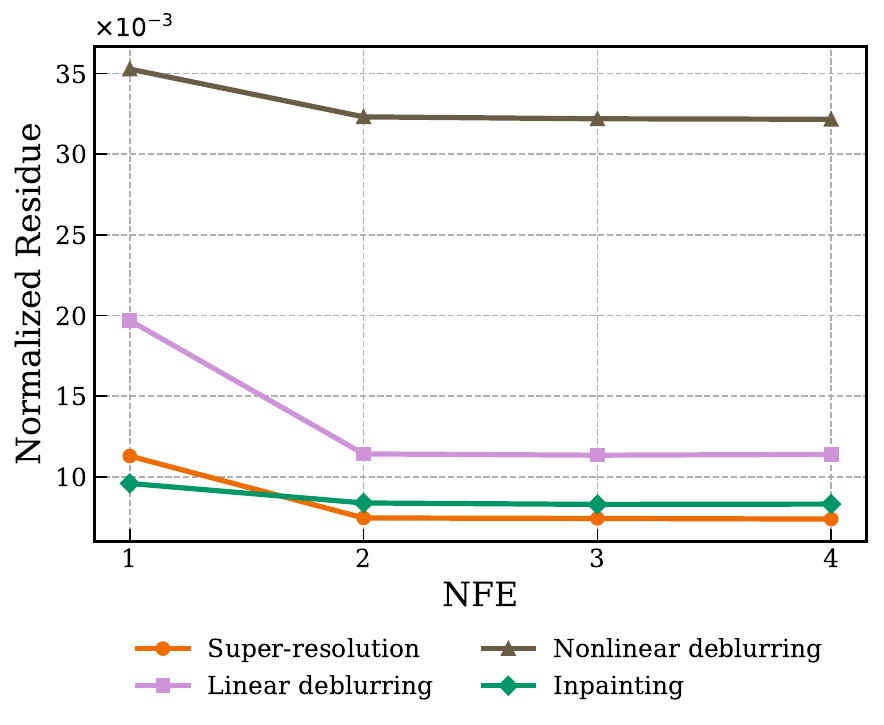}
}
\hfill
\subfloat[Fashion-MNIST]{
    \includegraphics[width=0.48\textwidth]{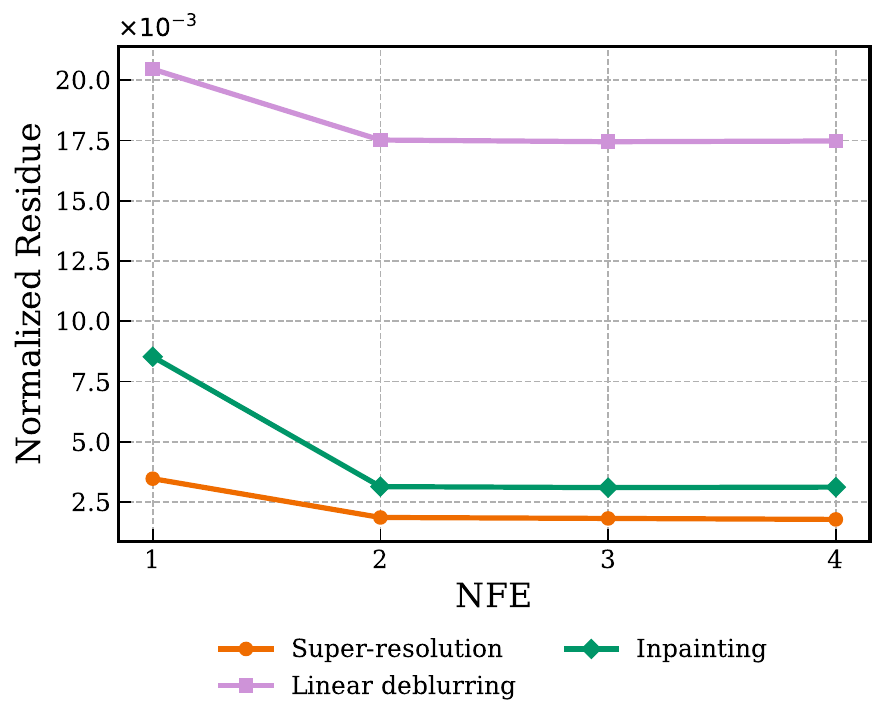}
}

\caption{Normalized measurement residue as a function of NFEs for MACS on the (a) LSUN Bedroom and (b) Fashion-MNIST datasets. MACS achieves most of its reduction in measurement residue within the first two steps across all inverse problems, after which the residue plateaus.}

\label{fig:residue-both}
\end{figure*}

\subsection{Evaluation}

We compare the proposed method with the standard few-step CM sampler~\cite{song2023consistency} and several fast ODE-based samplers, including 2-step Euler~\cite{kerras2022}, 8-step DPM-Solver~\cite{dpm-solver}, and 40-step Heun~\cite{kerras2022}. All baselines use the same CM backbone, time parameterization, and post-processing. The CM predicts a denoised estimate that is converted into the vector field required by each ODE solver, ensuring that differences in performance arise solely from the sampling strategy.

The hyperparameter $\gamma$ is chosen empirically for each task. On LSUN Bedroom, we use $\gamma=0.5$ for nonlinear deblurring, $\gamma=0.4$ for linear deblurring, and $\gamma=0.8$ for both inpainting and super-resolution. On Fashion-MNIST, a fixed value of $\gamma=0.15$ is used for all tasks. Each baseline is evaluated using the smallest number of sampling steps that achieves its best performance, while our method consistently performs best with only two sampling steps, with additional NFEs providing no measurable benefit.

Performance is evaluated using both reconstruction and perceptual metrics, including PSNR, SSIM~\cite{ssim}, KID ($\times10^3$)~\cite{kid}, and FID~\cite{fid}.

\subsection{Results}

Quantitative results for super-resolution, linear deblurring, and inpainting are reported in Tables~\ref{tab:results-LSUN-bedroom} and~\ref{tab:results-Fashion-MNIST}, while nonlinear deblurring results are shown in Table~\ref{tab:results-NonlinearDeblur}. Qualitative comparisons are presented in Figs.~\ref{fig:comparison-lsun} and~\ref{fig:comparison-Fashion-mnist}. Across both datasets, MACS consistently achieves the best perceptual quality, obtaining lower FID and KID scores than all baselines while maintaining competitive PSNR and SSIM values.

On Fashion-MNIST, MACS outperforms all competing methods across most metrics and inverse tasks. On LSUN Bedroom, it consistently produces the best perceptual results, with only a slight reduction in pixel-level metrics compared with the strongest baselines. 

As shown in Fig.~\ref{fig:residue-both}, the normalized measurement residue decreases rapidly within the first two sampling steps, after which additional NFEs provide negligible improvement, confirming that MACS effectively converges in only two iterations.

Overall, the proposed method achieves high-quality reconstructions without retraining or modifying the CM denoiser and generalizes across different inverse problems and conditioning strategies. The minor decrease in PSNR and SSIM observed in some settings is consistent with the well-known perception--distortion trade-off, where perceptually realistic reconstructions may sacrifice slight pixel-wise accuracy~\cite{saharia2022palette,blau2018perception,FreirichMM21,cohen2024lookstogood}.

\section{Conclusion}
\label{sec:conclusion}

We introduced MACS, a measurement-aware consistency sampling method that improves reconstruction quality for both linear and nonlinear inverse problems while preserving the fast sampling efficiency of consistency models. Experiments on the Fashion-MNIST and LSUN Bedroom datasets demonstrate consistent gains in perceptual quality with competitive pixel-level performance. Owing to its adaptability across different inverse problems and CM architectures, MACS serves as a plug-and-play enhancement for existing CM-based inverse solvers.

\bibliographystyle{IEEEtran}
\bibliography{ref}

\end{document}